\documentclass[conference]{IEEEtran}
\IEEEoverridecommandlockouts
\usepackage{cite}
\usepackage{graphicx}
\usepackage{comment}
\usepackage{balance}
\usepackage{acronym}
\usepackage{amsmath}
\usepackage{tabularx}
\usepackage[usenames,dvipsnames]{xcolor}
\usepackage[hang]{subfigure}
\usepackage{booktabs}
\usepackage{tikz}
\usetikzlibrary{positioning}
\usepackage{caption}
\usetikzlibrary{shapes.multipart}
\usepackage{amsfonts}
\usepackage{enumitem,bbm}
\usepackage{wrapfig}

\newcommand{\norm}[1]{\left\lVert#1\right\rVert}

\acrodef{STFT}{short-time Fourier transform}
\acrodef{ISTFT}{inverse short-time Fourier transform}
\acrodef{BSS}{blind source separation}
\acrodef{DOA}{direction of arrival}
\acrodef{DC}{deep clustering}
\acrodef{DPRNN}{dual-path recurrent neural network}
\acrodef{TF}{time-frequency}
\acrodef{TCN}{temporal convolutional network}
\acrodef{ATF}{acoustic transfer function}
\acrodef{SI-SDR}{scale-invariant signal-to-distortion ratio}
\acrodef{MSE}{mean square error}
\acrodef{DFT}{discrete Fourier transform }
\acrodef{SIR}{signal-to-interference ratio}
\acrodef{OVA}{overlap-and-add}
\acrodef{SDR}{signal-to-distortion ratio}
\acrodef{BLSTM}{Bidirectional Long Short-Term Memory}
\acrodef{SOTA}{state-of-the-art}
\acrodef{RI}{Real-Imaginary}
\acrodef{RIR}{room impulse response}
\acrodef{SNR}{signal-to-noise ratio}
\acrodef{RNN}{Recurrent Neural Networks}
\acrodef{FLOP}{floating point operation}
\acrodef{WPE}{weighted prediction error}
\acrodef{CASA}{computational simultaneous grouping scene analysis}
\acrodef{FC}{fully connected}
\acrodef{E2E}{end-to-end}
\acrodef{SDR}{signal to distortion ration}
\acrodef{SIR}{signal-to-interference ratio}
\acrodef{STOI}{short-time objective intelligibility}
\acrodef{PESQ}{perceptual evaluation of speech quality}
\acrodef{CNN}{convolutional neural network}
\tikzset{
  block/.style={
    rectangle,
    rounded corners,
    draw=black,
    very thick,
    minimum height=3em,
    minimum width=3em
  }
}
\tikzset{
  input/.style={
    circle,
    draw=black,
    very thick,
    minimum size=2em
  }
}
\tikzset{
  output/.style={
    circle,
    draw=black,
    very thick,
    minimum size=2em
  }
}

\ifCLASSINFOpdf

\else

\fi

\title{A two-stage  speaker extraction algorithm under adverse acoustic conditions using a single-microphone
\thanks{This project has received funding from the European Union’s Horizon 2020
Research and Innovation Programme under Grant Agreement No.~871245.}}

\author{\IEEEauthorblockN{Aviad Eisenberg}
\IEEEauthorblockA{Bar-Ilan University, OriginAI}
\and
\IEEEauthorblockN{Sharon Gannot}
\IEEEauthorblockA{Bar-Ilan University}
\and
\IEEEauthorblockN{ Shlomo E. Chazan}
\IEEEauthorblockA{OriginAI, Bar-Ilan University}
}

\begin{document}

\maketitle

\begin{abstract}
In this work, we present a two-stage method for speaker extraction under reverberant and noisy conditions. Given a reference signal of the desired speaker, the clean, but the still reverberant, desired speaker is first extracted from the noisy-mixed signal. In the second stage, the extracted signal is further enhanced by joint dereverberation and residual noise and interference reduction. The proposed architecture comprises two sub-networks, one for the extraction task and the second for the dereverberation task. We present a training strategy for this architecture and show that the performance of the proposed method is on par with other \ac{SOTA} methods when applied to the WHAMR! dataset. Furthermore, we present a new dataset with more realistic adverse acoustic conditions and show that our method outperforms the competing methods when applied to this dataset as well. 

\end{abstract}


\IEEEpeerreviewmaketitle

\section{Introduction}
Extracting a desired speaker from a mixture of overlapping speakers using  only a single microphone is a cumbersome task, particularly in noisy and reverberant environments. In this paper, we address this challenge by focusing on the extraction of a single participant from a mixture of two speakers acquired by a single microphone, given a prerecorded utterance of the speaker to be extracted. 


There has been significant progress in the single-microphone \ac{BSS} domain in the past years. The Conv-Tasnet \cite{luo2019conv} and the \ac{DPRNN} \cite{luo2020dual}, are both applied in the time domain with similar encoder-masking-decoder architecture. 
Other works that followed this approach were presented
\cite{subakan2021attention,wang2021end,lutati2022sepit,chazan2021single,nachmani2020voice,chen2020dual,tzinis2022compute,zeghidour2021wavesplit}, demonstrating a considerable improvement in the separation results. The SepFormer was introduced in \cite{subakan2021attention} leveraging the benefits of the attention layers, which led to a significant improvement in performance and to \ac{SOTA} results. 
An efficient \ac{CNN}-based model, denoted Sudo rm-rf, was presented in \cite{tzinis2022compute} and demonstrated high separation capabilities.
Most of the above-mentioned \ac{BSS} models were trained and tested on clean and anechoic mixtures. Such acoustic conditions can hardly be met in reality. Several algorithms~\cite{chazan2021single,subakan2021attention,tzinis2022compute,zeghidour2021wavesplit} were also trained on reverberant data without any changes in their architecture. Cord-Landwehr et al.~showed in \cite{cord2022monaural} that despite the significant improvement achieved in clean conditions, only marginal improvements can be obtained in realistic reverberant and noisy conditions. 

Given a reference signal of the desired speaker turns the \ac{BSS} problem into an extraction problem, in which the permutation problem is alleviated. 
The SpeakerBeam algorithm, introduced in \cite{vzmolikova2019speakerbeam}, estimates a mask for the desired speaker in the spectral domain using the spectrum of the reference signal. While magnitude-domain processing might be sufficient in clean and anechoic conditions, it might be insufficient in noisy and reverberant conditions. In \cite{delcroix2020improving}, this model was improved by using the time-domain signal, as it allows the exploitation of the entire  signal information. A similar approach was presented in \cite{xu2019time}, where the i-vector \cite{dehak2010front} of the reference signal was used as the embedding of the desired speaker. In \cite{xu2020spex}, a multi-task training procedure was proposed in which a speaker classification task is carried out in parallel for improving the embedding of the desired speaker.

Time domain processing, despite the above advantages, ignores the time-frequency patterns typical to speech signals. In our prior work, \cite{eisenberg2022single}, a fully convolutional Siamse-Unet architecture was proposed. The algorithm is applied in the \ac{STFT} domain to the Real-Imaginary (RI) representation of the signals {while the loss is applied in the time-domain}, exploiting the entire signal, on the one hand, and leveraging its spectral patterns, on the other hand. Yet, the performance of this approach is insufficient in adverse acoustic conditions.

\begin{figure*}[th!]
\centering

\tikzset{every picture/.style={line width=0.75pt}} 

\begin{tikzpicture}[x=0.75pt,y=0.75pt,yscale=-1,xscale=1]

\draw   (44.6,71.31) -- (93.42,91.28) -- (93.38,103.68) -- (44.44,123.48) -- cycle ;
\draw   (207.95,122.76) -- (158.62,103.13) -- (158.58,91.41) -- (207.8,71.61) -- cycle ;
\draw   (125.1,161.85) -- (130.1,161.85) -- (130.1,191.35) -- (125.1,191.35) -- cycle ;
\draw    (138.8,178) -- (354.8,178.43) -- (354.2,155) ;
\draw    (127.2,154.8) -- (128.27,121.67) ;
\draw [shift={(128.33,119.67)}, rotate = 91.85] [color={rgb, 255:red, 0; green, 0; blue, 0 }  ][line width=0.75]    (10.93,-3.29) .. controls (6.95,-1.4) and (3.31,-0.3) .. (0,0) .. controls (3.31,0.3) and (6.95,1.4) .. (10.93,3.29)   ;
\draw   (112.8,82.66) -- (117.8,82.66) -- (117.8,112.16) -- (112.8,112.16) -- cycle ;
\draw   (117.8,82.66) -- (122.8,82.66) -- (122.8,112.16) -- (117.8,112.16) -- cycle ;
\draw   (122.8,82.66) -- (127.8,82.66) -- (127.8,112.16) -- (122.8,112.16) -- cycle ;
\draw   (127.8,82.66) -- (132.8,82.66) -- (132.8,112.16) -- (127.8,112.16) -- cycle ;
\draw   (132.8,82.66) -- (137.8,82.66) -- (137.8,112.16) -- (132.8,112.16) -- cycle ;
\draw   (137.8,82.66) -- (142.8,82.66) -- (142.8,112.16) -- (137.8,112.16) -- cycle ;
\draw  [dash pattern={on 4.5pt off 4.5pt}]  (234,97.6) -- (233.4,62.6) -- (233.4,47.3) -- (20.6,47) -- (20.6,74.6) ;
\draw [shift={(20.6,76.6)}, rotate = 270] [color={rgb, 255:red, 0; green, 0; blue, 0 }  ][line width=0.75]    (10.93,-3.29) .. controls (6.95,-1.4) and (3.31,-0.3) .. (0,0) .. controls (3.31,0.3) and (6.95,1.4) .. (10.93,3.29)   ;
\draw   (19.6,218.37) .. controls (19.61,223.04) and (21.95,225.37) .. (26.62,225.36) -- (117.42,225.1) .. controls (124.09,225.08) and (127.43,227.4) .. (127.44,232.07) .. controls (127.43,227.4) and (130.75,225.06) .. (137.42,225.05)(134.42,225.05) -- (228.22,224.79) .. controls (232.89,224.78) and (235.21,222.44) .. (235.2,217.77) ;
\draw   (239.2,218.07) .. controls (239.23,222.74) and (241.57,225.06) .. (246.24,225.03) -- (336.84,224.46) .. controls (343.51,224.42) and (346.86,226.73) .. (346.89,231.4) .. controls (346.86,226.73) and (350.17,224.38) .. (356.84,224.33)(353.84,224.35) -- (447.44,223.76) .. controls (452.11,223.73) and (454.43,221.39) .. (454.4,216.72) ;
\draw   (275,72.31) -- (323.82,92.28) -- (323.78,104.68) -- (274.84,124.48) -- cycle ;
\draw   (434.35,123.76) -- (385.02,104.13) -- (384.98,92.41) -- (434.2,72.61) -- cycle ;
\draw   (339.2,83.66) -- (344.2,83.66) -- (344.2,113.16) -- (339.2,113.16) -- cycle ;
\draw   (344.2,83.66) -- (349.2,83.66) -- (349.2,113.16) -- (344.2,113.16) -- cycle ;
\draw   (349.2,83.66) -- (354.2,83.66) -- (354.2,113.16) -- (349.2,113.16) -- cycle ;
\draw   (354.2,83.66) -- (359.2,83.66) -- (359.2,113.16) -- (354.2,113.16) -- cycle ;
\draw   (359.2,83.66) -- (364.2,83.66) -- (364.2,113.16) -- (359.2,113.16) -- cycle ;
\draw   (364.2,83.66) -- (369.2,83.66) -- (369.2,113.16) -- (364.2,113.16) -- cycle ;
\draw   (43.6,152.4) -- (92.42,172.37) -- (92.38,184.77) -- (43.44,204.57) -- cycle ;
\draw    (208,99) -- (273,99) ;
\draw [shift={(275,99)}, rotate = 180] [color={rgb, 255:red, 0; green, 0; blue, 0 }  ][line width=0.75]    (10.93,-3.29) .. controls (6.95,-1.4) and (3.31,-0.3) .. (0,0) .. controls (3.31,0.3) and (6.95,1.4) .. (10.93,3.29)   ;
\draw    (127.2,154.8) -- (138.34,122.89) ;
\draw [shift={(139,121)}, rotate = 109.24] [color={rgb, 255:red, 0; green, 0; blue, 0 }  ][line width=0.75]    (10.93,-3.29) .. controls (6.95,-1.4) and (3.31,-0.3) .. (0,0) .. controls (3.31,0.3) and (6.95,1.4) .. (10.93,3.29)   ;
\draw    (127.2,154.8) -- (117.57,122.25) ;
\draw [shift={(117,120.33)}, rotate = 73.51] [color={rgb, 255:red, 0; green, 0; blue, 0 }  ][line width=0.75]    (10.93,-3.29) .. controls (6.95,-1.4) and (3.31,-0.3) .. (0,0) .. controls (3.31,0.3) and (6.95,1.4) .. (10.93,3.29)   ;
\draw    (354,156) -- (355.07,122.87) ;
\draw [shift={(355.13,120.87)}, rotate = 91.85] [color={rgb, 255:red, 0; green, 0; blue, 0 }  ][line width=0.75]    (10.93,-3.29) .. controls (6.95,-1.4) and (3.31,-0.3) .. (0,0) .. controls (3.31,0.3) and (6.95,1.4) .. (10.93,3.29)   ;
\draw    (354,156) -- (365.14,124.09) ;
\draw [shift={(365.8,122.2)}, rotate = 109.24] [color={rgb, 255:red, 0; green, 0; blue, 0 }  ][line width=0.75]    (10.93,-3.29) .. controls (6.95,-1.4) and (3.31,-0.3) .. (0,0) .. controls (3.31,0.3) and (6.95,1.4) .. (10.93,3.29)   ;
\draw    (354,156) -- (344.37,123.45) ;
\draw [shift={(343.8,121.53)}, rotate = 73.51] [color={rgb, 255:red, 0; green, 0; blue, 0 }  ][line width=0.75]    (10.93,-3.29) .. controls (6.95,-1.4) and (3.31,-0.3) .. (0,0) .. controls (3.31,0.3) and (6.95,1.4) .. (10.93,3.29)   ;

\draw (15,86.96) node [anchor=north west][inner sep=0.75pt]    {$x$};
\draw (453,90.21) node [anchor=north west][inner sep=0.75pt]    {$\hat{s}_{d}$};
\draw (12.2,162.13) node [anchor=north west][inner sep=0.75pt]    {$\tilde{s}_{d}^{\text{ref}}$};
\draw (237.2,61.15) node [anchor=north west][inner sep=0.75pt]    {$\widetilde{\hat{s}}_{d}^{( L-1)}$};
\draw (118.2,194.17) node [anchor=north west][inner sep=0.75pt]    {$E_{d}^{\text{ref}}$};
\draw (111,233.56) node [anchor=north west][inner sep=0.75pt]   [align=left] {Stage 1};
\draw (328,231.29) node [anchor=north west][inner sep=0.75pt]   [align=left] {Stage 2};
\draw (116.8,9.95) node [anchor=north west][inner sep=0.75pt]    {$\widetilde{\hat{s}}_{d}^{( l-1)}$};
\draw (95.93,165.45) node [anchor=north west][inner sep=0.75pt]  [rotate=-359.76]  {${\displaystyle \sum }$};

\end{tikzpicture}
\caption{Block diagram of the proposed two-stage architecture. {In the first iteration, the model takes the given observations as input. For subsequent iterations, the output of the previous iteration is used as input instead of the mixture. The network is using skip connections between the mixture encoder and decoder (not shown explicitly in the diagram). No skip connections from the reference encoder are implemented. The two encoders share the same weights. The arrows denote the element-wise multiplication of the reference embedding with the embedding of each frame. Only the output of the final iteration is used as input to the second stage.}}
\label{fig:block_diagram}
\end{figure*}
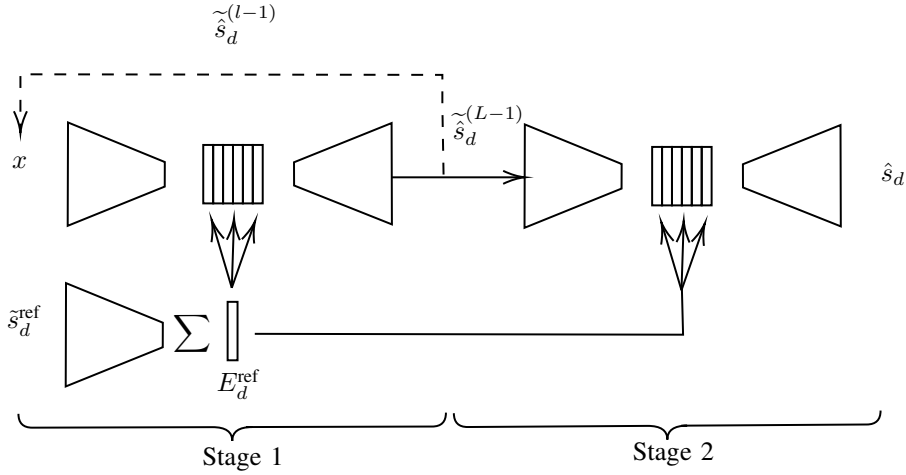


In the current contribution, we present a two-stage algorithm to extract a desired speaker from a mixture of two signals under reverberant and noisy conditions. We split the extraction task into two stages. In the first stage, given the noisy and reverberant mixture and the reference signals, a Siamse-Unet architecture is applied to extract the \emph{reverberant} desired speaker.
{The encoders used for both the mixture and the reference signals are identical, thus the resulting outputs have matching dimensions. While the mixture encoder preserves the frame dimensions, which is essential for the mixture processing, the reference encoder aims to exclusively represent the desired speaker's identity while ignoring the content of the utterance. To achieve this outcome, we average the reference embedding over the frame dimension. The reference embedding vector is finally multiplied with each of the frames in the mixture embedding. The outcome of this multiplication is used as an input to the decoder, which in turn extracts the reverberant desired speaker.}
We show that training this stage in an iterative manner is beneficial. 

In the second stage, an additional Unet model is applied to dereverberate and enhance the output of the first stage. Similarly, the encoder output preserves the {frame size }
of its input signal. The resulting embedding is multiplied by the embedding of the reference from the first stage. The second decoder is finally applied to extract the desired \emph{clean and dereverberated} signal.  

Furthermore, in this paper, we introduce a new simulated dataset with more realistic conditions than the WHAMR! dataset, and show that our model outperforms  other \ac{SOTA} models on both the WHAMR! dataset and the new, more challenging, dataset.

\section{Problem Formulation}
The signal $x(t)$, captured by a single microphone, is a combination of $Q$ concurrent speakers, represented by:
\begin{equation}
    x(t) = \sum^Q_{q=1} \{s_q\ast {h}_q\}(t) + v(t) \quad {t=0,1,\ldots,T-1}
    \label{eq:prob_time}
\end{equation}
where $s_q(t)$ is the signal of the $q$th speaker, ${h}_q(t)$ is the \ac{RIR} between the $q$th speaker position and the microphone position, and $v(t)$ is an additive noise. In a noise-free, non-reverberant environment, ${h}_q(t)$ {is dominated by the first arrival}, and $v(t)=0$ for all $q$.

In the \ac{STFT} domain, the microphone signal can be approximately expressed as:
\begin{equation}
    x(n,k) = \sum^Q_{q=1} s_q(n,k)  h_q(n,k) + v(n,k)
    \label{eq:prob_freq}
\end{equation}
where $n=0,1,\ldots,N-1$ and $k=0,1,\ldots,K-1$ represent the time-frame and frequency-bin indexes, respectively, and $N$ and $K$ are the total number of time-frames and frequency bands, respectively.

This paper focuses on the case where there are only two concurrent speakers, namely $Q=2$, referred to as the desired speaker $s_d(n,k)$ and the interference speaker $s_i(n,k)$. 
The reverberant desired signal is defined as $\tilde{s}_d(n,k) = s_d(n,k)  h_d(n,k)$.
The reference signal is denoted  $s_d^{\text{ref}}(n,k)$. We aim at the extraction of the desired speaker signal, $\hat{s}_d(n,k)$, using the mixed signal $x(n,k)$, and a reverberant reference signal, $\tilde{s}_d^{\text{ref}}(n,k) = s_d^{\text{ref}}(n,k) h_d^{\text{ref}}(n,k)$.

\section{Proposed Model}

\subsection{Architecture and Training Procedure}
Our model is composed of two sub-stages. The first is a Simase-Unet, which consists of three parts: two  encoders and a decoder. We share weights between the encoders to encourage joint embeddings of both the mixture and the reference signals in the same latent space.
{The encoder architecture consists of several convolution layers followed by two-dimensional batch normalization and a `Relu' function (similar to the one introduced in \cite{eisenberg2022single}). Next, we combine the dimensions of the channels and frequencies and employ a fully-connected layer to reduce the dimensions. After this step, we apply a single transformer-encoder layer. The decoder architecture consists of six transformer-encoder layers, followed by \ac{FC} layer to restore the original dimension. Then transpose-convolution layers are employed to adapt to the convolution layers in the encoder, enabling the application of skip connections as required. A transformer-encoder layer is subsequently applied after all the steps mentioned above.}
We repeat the first stage several times to further enhance the extraction process. In the first iteration, the mixture signal is processed, while in the subsequent iterations, the separated (but still reverberant) signals from the previous iteration are processed. 
Formally, the process can be expressed as:
\begin{equation*}
    \textrm{Input}^{(\ell)} =
    \begin{cases}
      x(n,k) &   \ell=0 \\
      \hat{\tilde{s}}_d^{(\ell-1)}(n,k)  & \ell>0
    \end{cases}
\end{equation*}
where $\ell = 0,...,L-1$ is the iteration index.
By repeating this process for $L$ iterations, we obtain $L$ estimates of $\tilde{s}_{d}(n,k)$, which are all used to train the entire model.

The second stage of the model uses the same architecture as the first stage. Our empirical results showed that using the {reveberant} reference signal in the second phase can improve the results. Rather than passing the reference signal again through an encoder, we can simply use the learned embedding vector from the first stage.

Alternative ways for integrating the information from the reference signal are described in \cite{vzmolikova2019speakerbeam}, including concatenation, addition, and multiplication, the latter achieving the best results. {To obtain a single vector that represents the speaker's identity, we average across the frame dimensions of the reference embedding, thus ignoring the temporal information and emphasizing the speaker's identity. The final embedding vector is denoted $E^{\text{ref}}_d$.} Unlike \cite{eisenberg2022single},  {in the Unet architecture, skip connections are only implemented from the mixture encoder and not from the reference encoder}. Instead, we only use the output of the last layer of the reference encoder in the bottleneck stage. While most {single microphone DNN-based} algorithms apply a masking operation to the mixture signal, the proposed scheme is trained to directly estimate the \ac{TF} representation of the target source.

The two sub-stages are trained together in an end-to-end manner, while the first stage feeds the second phase with an estimate of the last iteration of the first stage and the reference embedding. A block diagram of the entire model is shown in Fig.~\ref{fig:block_diagram}.

\subsection{Features}
In this work, we adopted the \ac{RI} components of the \ac{STFT} as both the input features of the model and its output. The model is trained with the \ac{SI-SDR} loss function, which is sensitive to phase distortion.  Using the \ac{RI} features may alleviate such problems (see discussion in \cite{eisenberg2022single}).  

\subsection{Objectives}
As mentioned above, we  use the \ac{SI-SDR} loss function to train our model. The loss is formulated as
\begin{equation}
\text{SI-SDR}\left( s,\hat{s} \right)= 10 \log_{10} \left( \frac{\norm{\frac{\langle {\hat{s},s} \rangle}{\langle {s,s} \rangle} s}^2}{\norm{\frac{\langle {\hat{s},s} \rangle}{\langle {s,s} \rangle} s-\hat{s}}^2} \right).
\end{equation}
The model is trained using all output signals, {namely, $\hat{s}_d$ and  $\hat{\tilde{s}}_d^{(\ell)}, \ell = 0 , \ldots , L-1$:} 
\begin{equation}
\mathcal{L}_{\text{SISDR}_d} = \sum_{\ell=0}^{L-1} \ac{SI-SDR}\left( \tilde{s}_{d}, \hat{\tilde{s}}_d^{(\ell)} \right) + \ac{SI-SDR}\left( s_{d},\hat{s}_d \right).
\label{eq:L_sisdr}
\end{equation}
For the extraction task to be successful, the network must be able to learn a unique embedding for each speaker to prevent errors in identifying the correct speaker. To achieve this goal, an additional, triplet loss function, was implemented:
\begin{equation}
{\text{TRIPLET}}(a,p,n) = \max( \text{cd}(a,p) - \text{cd}(a,n) + m, 0)
\end{equation}
where $a$ is the anchor input, $p$ is the positive input and $n$ is the negative input, with $p$ closer to $a$ than $n$. The function $\textrm{cd}(\cdot)$ is the cosine distance and $m$ is a margin hyperparameter. The triplet loss function encourages the distance between the anchor and the positive inputs to be smaller than the distance between the anchor and negative inputs, by a margin of at least $m$.
In our case, we would like the embedding of the reverberant reference to be as close as possible to the embedding of the output of the first phase (namely, the estimated desired and reverberant speaker), and as far as possible from the embedding of the reference of the second speaker. In explicit terms:
\begin{equation}
\mathcal{L}_{\text{TRIPLET}_d} = 
{\text{TRIPLET}}(E_{\hat{\tilde{s}}_d},E^{\text{ref}}_d,\overline{E_d^{\text{ref}}} )
\end{equation}
where $E_{\hat{\tilde{s}}_d}$ is obtained by passing $\hat{\tilde{s}}_{d_{L-1}}$ through the encoder of stage 1 and $\overline{E_d^{\text{ref}}}$ is the embedding of the reference of the interference signal.

During training, we encountered a convergence problem when using both loss {functions} simultaneously. To address this issue, we implemented a warm-up training procedure in which the network is initially trained using only the SI-SDR loss, and the triplet loss is added at a later stage in the training process. This approach successfully resolved the convergence issues.

In an effort to improve the training process, we alternated the desired and interference signals within each training batch, while maintaining consistency in the mixture employed. {That is, inserting the mixture signal with the reference signal of one of the speakers and then repeating the process with the reference of the other speaker in the same batch, and summing the losses for both speakers.} 
In short, the overall loss function takes the following form:
\begin{multline}
\mathcal{L} =  (\mathcal{L}_{\text{SISDR}_d} + \mathcal{L}_{\text{SISDR}_i})/2 \;  +   \\ \alpha \cdot \mathbbm{1}_{\textrm{warm-up}}  \cdot (\mathcal{L}_{\text{TRIPLET}_d} + \mathcal{L}_{\text{TRIPLET}_i} )/2
\label{eq:L_triplet}
\end{multline}
where $\alpha$ represents a hyperparameter, and the indicator function $\mathbbm{1}_{\textrm{warm-up}}$ determines the point at which the triplet objective function should be taken into consideration in the training process.

\section{Experimental Study}

\subsection{Datasets}
We used the WHAMR! dataset to train our model. This dataset is created by taking the WSJ0-2Mix dataset \cite{hershey2016deep} and modifying it by incorporating environmental noise from the WHAM dataset \cite{wichern2019wham} and reverberation. 
To adapt the dataset to the extraction task, we modified it in the following manner. For each speaker included in the mixture, we selected a different utterance and convolve it with the same room impulse response (RIR) used to generate the mixed signal, namely $h_d^{\text{ref}} = h_d$. This procedure reflects the fact that in a typical conversation, segments in which only a single speaker is active can always be found. However, it is implicitly assumed that the scenario is static, hence that the RIR does not significantly change during the entire conversation. 

We note that, according to our tests, the reverberation level in the WHAMR! dataset does not exceed 600 milliseconds, in contradiction to the reported reverberation level, which is in the range of $[0.2, 1]$.\footnote{Due to space constraints, we will not give a detailed analysis of the dataset in the current contribution.} 

The dataset includes 20,000 signals for training, 5,000 for validation, and 300 for the test phase, and it uses the `min' and `8k' {sampling rate} configuration. {(With `min' setting the longer target is truncated to match the length of the shorter target.)}

In addition to WHAMR!, we generated a new dataset for the purpose of enriching the data. This is equivalent to \emph{dynamic mixing} training, which randomly generates the mixture from the existing speakers during training. We also took speakers from the WSJ0 corpus, along with noise from the WHAM and the reverberation generated from an RIR generator \cite{habets2006room} with parameters listed in Table~\ref{table:reverb_parameters}. 

During training, each signal is truncated to a variable length between 2 to 5 seconds. Since we are using a Siamese architecture, the mixture and the reference signal must have the same length. If the reference signal is longer, it will be truncated, and if it is shorter, it will be duplicated until it is the same length as the mixture.

\begin{table}[t]
\addtolength{\belowcaptionskip}{6pt}
\caption{Noisy reverberant data specification.}
\label{table:noisy_data}
\centering
\resizebox{0.8\columnwidth}{!}{
\begin{tabular}{@{}lll@{}}
\toprule
                        & $H_x$         & $U[4,8]$                                 \\
Room dim.~[m]                & $H_y$        & $U[4,8]$                                  \\
                        & $H_z$        & $U[2.5,3]$                                               \\ \midrule
Reverb. time~[sec]             &     $T_{60}$      & $U[0.2,0.6]$                            \\ \midrule
                        & $x$        & $\frac{H_x}{2}+{U}[-0.5,0.5]$ \\
Mic. Pos. [m]           &$y$         & $\frac{H_y}{2}+{U}[-0.5,0.5]$ \\
                        & $z$        & 1.5                                         \\ \midrule

Sources Pos. [$^\circ$] & $\theta$ & \emph{U}{[}0,180{]}                                \\ \midrule
Sources Distance [m]                   &          & $1+{U}[-0.5,0.5]$                        
            \\ \bottomrule
\end{tabular}}
\label{table:reverb_parameters}
\end{table}

\subsection{Algorithm Settings}
The frame-size of the \ac{STFT} is  256 samples with $50\%$ overlap.  Due to the symmetry of the \ac{DFT} only the first half of the frequency bins are used. The value of $\alpha$ was empirically set to 2, emphasizing the triplet loss due to the significant difference in scales between the two objective functions. The triplet loss margin was set to $m=0.5$.

The number of iterations for the first phase was chosen as $L=2$, because there was minimal improvement when increasing the number from 2 to 3 iterations, while a noticeable improvement was observed between 2 iterations to no iterations, $L=1$.

In the training procedure, we used the Adam optimizer~\cite{kingma2015adam}. The learning rate was set to 0.001 and the training batch size to 6. The weights are randomly initialized, and the lengths of the signals were randomly changed at each batch.

\subsection{Evaluation Measures}
To evaluate the proposed algorithm we use five evaluation measures: \ac{SI-SDR}, \ac{SIR}, \ac{SDR}, \ac{STOI}, and \ac{PESQ}. While the first three are used as a measurement of the quality of the speaker separation, the last two give an indication of the audio intelligibility and quality. 

The proposed algorithm is compared to the current \ac{SOTA} separation methods, i.e., the Sepformer \cite{subakan2021attention} and the Sudo rm-rf \cite{tzinis2022compute}. These are time-domain blind source separation masking-based methods. We decided to compare our method with separation methods rather than extraction methods since these are the most effective methods in the field.

\subsection{Results}
The results for the WHAMR! dataset are depicted in Table \ref{table:whamr}. Our model achieves an SI-SDR of 9.67~dB, SDR of 10.88~dB, and SIR of 24.2~dB. It is evident that our proposed method outperforms the \ac{SOTA} methods in almost all measures. In addition, the method also achieves the best scores for the intelligibility measure (STOI) and the quality measure (PESQ), with scores 92\% and 2.72, respectively.

The new dataset imposes a greater challenge on the extraction algorithm, as evidenced by the lower scores in Table~\ref{table:mydata} for all measures, compared to the scores obtained on the WHAMR! dataset, as reported in Table~\ref{table:whamr}. While the absolute separation results obtained for the new dataset are lower, the improvement in terms of SI-SDR is 14.2~dB, which is very high and significantly outperforms the competing methods. The intelligibility results (90.2\%) are on par with the results obtained for the WHAMR! dataset.

\begin{table}[htbp]
\caption{Results for WHAMR! dataset}
\begin{center}
\begin{tabular}{@{}lcccccc@{}}
\toprule
  Model & SI-SDR &  SDR &  SIR   &  STOI  & PESQ   \\
  \midrule
Unprocessed   & -3.84 & -0.59 & 0.19 & 65.3  & 1.51 \\ 
Sudo rm-rf\cite{tzinis2022compute}   & 8.13 & 10.7 &  23.7  & 90.2 & 2.5 \\

Sepformer\cite{subakan2021attention}   & 8.86 &  10 & \textbf{25}  & 91.3 & 2.57 \\

Proposed   & \textbf{9.67} & \textbf{10.88}  & 24.2  & \textbf{92} & \textbf{2.72} \\
\bottomrule
\end{tabular}
\end{center}
\label{table:whamr}
\end{table}

\begin{table}[htbp]
\caption{Results for the new dataset}
\begin{center}
\begin{tabular}{@{}lcccccc@{}}
\toprule
  Model & SI-SDR &  SDR &  SIR   &  STOI  & PESQ   \\
  \midrule
Unprocessed   & -7.99 & -0.79 & 0.12 & 52.5  & 1.54 \\ 
Sudo rm-rf\cite{tzinis2022compute}   & 1.7 & 3.46  & 15.8   & 69.9 & 2.1 \\

Sepformer\cite{subakan2021attention}   & 1.89 & 4.82  & 18.48  & 68.8 & 2.05 \\

Proposed   & \textbf{6.21} & \textbf{7.98}  &  \textbf{22.14} & \textbf{90.2} & \textbf{2.62} \\
\bottomrule
\end{tabular}
\end{center}
\label{table:mydata}
\end{table}

\subsection{Ablation Study}
We present an ablation study for our model. We examined four different configurations:

\begin{enumerate}

     \item One iteration in the first stage. The loss function for the desired source is given by:
\begin{equation}\mathcal{L}_{\text{SISDR}_d} = \ac{SI-SDR}\left( \tilde{s}_{d}, \hat{\tilde{s}}_d^{(L-1)} \right) + \ac{SI-SDR}\left( s_{d},\hat{s}_d \right)
\label{eq:ablation}
     \end{equation}
      with $L = 1$, and the overall loss is given by
      $\mathcal{L} =  (\mathcal{L}_{\text{SISDR}_d} + \mathcal{L}_{\text{SISDR}_i})/2 $.
      Triplet loss is not applied.
     \item Two iterations in the first stage. The SI-SDR loss is only applied to the final output $\hat{\tilde{s}}_{d}^{(L-1)}$, i.e. $L=2$ in \eqref{eq:ablation}. Triplet loss is not applied. 
     \item 
      The SI-SDR loss is applied to all intermediate results $\ell=0,\ldots, L-1$, as in \eqref{eq:L_sisdr}, with $L=2$.
     Triplet loss is not applied. 
     \item The full implementation of the proposed model with all its components active.
\end{enumerate}

Table~\ref{table:ablation} depicts the breakdown of the results for the WHAMR! and the new datasets. {It is evident that each additional component enhances the quality of the network output for both datasets. In total, the SI-SDR measure improved from 8.62~dB to 9.67~dB for the WHAMR! dataset and from 5.45~dB to 6.21~dB for the new dataset. Respectively, STOI improved from 90.4\% to 92\% for WHAMR!, and from 88\% to 90.2\% for the new dataset}

{Training the model to accurately identify the intended speaker from a mixture is challenging in speaker extraction, particularly in reverberant conditions and when the speakers have similar voices. This may result in the extraction of the incorrect speaker or a permutation between the output signals.}
To address this issue, the triplet loss was added. Our experiments showed that the addition of the triplet loss alleviated such permutation problems.







\begin{table}[htbp]
\caption{Ablation Study for all 4 configurations.}
\begin{center}
\begin{tabular}{@{}ccccccc@{}}

  \toprule

 &\multicolumn{2}{c}{WHAMR!} &\multicolumn{2}{c}{New}\\   \cmidrule(r){2-3} \cmidrule(r){4-5}
 Config.  & SI-SDR &  STOI &   SI-SDR &  STOI    \\
  \midrule
1)   & 8.62 & 90.4  & 5.45 & 88  \\ 
2)   & 9.13 &  91 &  5.71 &  88.9 \\

3)   & 9.26 &  91.8 &  6.02 &  \textbf{90.2} \\

4)  & \textbf{9.67} &  \textbf{92}  & \textbf{6.21} & \textbf{90.2} \\
\bottomrule
\end{tabular}
\vspace{1ex}

 \end{center}
\label{table:ablation}
\end{table}




\section{Conclusions}
We have proposed a two-stage approach for speaker extraction under reverberant conditions. The first stage separates the desired and yet reverberated speaker, while the second stage reduces reverberation and further enhances separation quality. Our results indicate that our model performs comparably or better than current state-of-the-art separation methods, with the added benefits of faster and more consistent training. Furthermore, an ablation study identifies the role of the various components in improving performance.

 \balance
\bibliographystyle{IEEEtran}
\bibliography{main}
\end{document}